\documentclass[a4paper,12pt,notitlepage,layout=twocolumn, superscriptaddress, manuscript=article, longbibliography,floatfix,authoryear,xcolor={table}]{article}%{achemso}

\usepackage[left=2cm, right=2cm]{geometry}
\usepackage{inputenc}
\usepackage[style=nature]{biblatex}
\usepackage{xcolor}

\usepackage{authblk}
\usepackage{graphicx} % Required for inserting images
\usepackage{siunitx,booktabs}
\usepackage{tabularx}
\usepackage[normalem]{ulem}
\usepackage[export]{adjustbox}
\usepackage{braket}
\usepackage{float}
\usepackage{tabularx}
\usepackage{amssymb}
\usepackage{makecell}
\usepackage{array, booktabs, makecell, multirow}
\usepackage[intlimits]{amsmath}
\usepackage{here}
\usepackage{csquotes}
\usepackage{tikz}
\usepackage{siunitx,booktabs}
\usepackage{makecell}
\usepackage{array,varwidth,lipsum}
\usepackage{booktabs}
\usepackage{siunitx}
\usepackage{color}
\usepackage{tabularray}
\usepackage{tcolorbox}

\usepackage{color, colortbl}
 \usepackage{url}
\definecolor{myorange}{RGB}{255,140,0}
\usepackage[colorlinks=true,linkcolor=Blue,citecolor=Blue,filecolor=Blue]{hyperref}
\definecolor{myblue}{rgb}{0.0, 0.18, 0.39}
\newcolumntype{d}[1]{D{.}{.}{#1}}
\newcommand\mc[1]{\multicolumn{1}{c}{#1}}
\hypersetup{
    colorlinks=true,
    urlcolor=black,
    citecolor=black,
    linkcolor=black}

\makeatletter
\newcommand{\setlabel}[1]{\edef\@currentlabel{#1}\label}
\makeatother

\newcommand{\thickcline}[1]{%
    \@thickcline #1\@nil%
}

\usepackage{etoolbox}           % <--
\renewcommand{\bfseries}{\fontseries{b}\selectfont} % <--
\robustify\bfseries             % <--
\newrobustcmd{\B}{\bfseries}

\newcommand{\thickhline}{%
    \noalign {\ifnum 0=`}\fi \hrule height 0.0pt
    \futurelet \reserved@a \@xhline
}
\newcolumntype{"}{@{\hskip\tabcolsep\vrule width 1.0pt\hskip\tabcolsep}}

\usepackage{booktabs}
\newcolumntype{C}[1]{>{\centering\let\newline\\\arraybackslash\hspace{0pt}}m{#1}}

\begin{document}\sloppy
\title{Magnetic anisotropy of $4f$ atoms on a WSe$_2$ monolayer: a DFT+U study}   
\author[1,2,*]{Johanna P. Carbone}
\author[1]{Gustav Bihlmayer}
\author[1]{, Stefan Bl\"{u}gel}

\affil[1]{Peter Gr\"unberg Institute and Institute for Advanced Simulation, Forschungszentrum J\"ulich and JARA, 52425 J\"ulich, Germany \looseness=-1}
\affil[2]{Institute of Theoretical Physics, Technical University of Vienna, 1040 Vienna, Austria}
\renewcommand\Authands{ } % avoid "and" before last author
\renewcommand\Affilfont{\itshape\small}
\date{}
%\maketitle
\let\oldmaketitle\maketitle
\let\maketitle\relax

\twocolumn[
  \begin{@twocolumnfalse}
\oldmaketitle
% Add your email
\begin{center}
 {$^{*}$johanna.carbone@tuwien.ac.at}   
\end{center}

\begin{abstract}
Inspired by recent advancements in the field of single-atom magnets, particularly those involving rare-earth (RE) elements, we present a theoretical exploration employing DFT+$U$ calculations to investigate the magnetic properties of selected $4f$ atoms, specifically Eu, Gd and Ho, on a monolayer of the transition-metal dichalcogenide WSe$_2$ in the 1H-phase. This study comparatively examines RE with diverse $4f$ orbital fillings and valence chemistry, aiming to understand how different coverage densities atop WSe$_2$ affect the magnetocrystalline anisotropy. We observe that RE elements lacking $5d$ occupation in the atomic limit exhibit larger magnetic anisotropy energies at high densities, while those with outer $5d$ electrons show larger anisotropies in dilute configurations. Additionally, even half-filled $4f$ shell atoms with small orbital magnetic moments can generate substantial energy barriers for magnetization rotation due to prominent orbital hybridizations with WSe$_2$. Open $4f$ shell atoms further enhance anisotropy barriers through spin-orbit coupling effects. This aspect is crucial for the experimental realization of stable magnetic information units.
\end{abstract}
\end{@twocolumnfalse}]

\section{Introduction}

The continuous demand of computational power and the continuous generation of increasing amount of data 
motivates novel computing paradigms and improved memory capacity in electronic devices. A
promising approach for both is the use of single magnetic atoms with large magnetic anisotropies deposited on
2D-materials or surfaces~\cite{donati2021perspective,khajetoorians2016toward,rau2014reaching,gambardella2003giant,brune2009magnetism}. These magnetic atoms can act as information carriers, with their magnetization aligned in specific preferred directions, energetically separated from other orientations. This property can also give rise to two-level quantum systems \cite{willke2019tuning,chen2023harnessing}.

In this context, rare-earth (RE) atoms on 2D materials offer intriguing possibilities for stable magnetic units in classical and quantum applications, serving as bits and qubits \cite{sessoli2017single,natterer2017reading,natterer2018thermal,reale2023erbium,PhysRevB.108.L180408}. This is attributed to their significant spin and orbital magnetic moments arising from the strongly localized $4f$ electrons. The selection of RE atoms as a magnetic source is indeed justified by the limited hybridization effects of the $4f$ electrons, ensuring the generated magnetization is protected and rigid. 
In fact, the essential requirement for utilizing magnetic atoms as information carriers is a stable magnetization, both in magnitude and orientation, safeguarded against reversal due to thermal or quantum fluctuations as well as environmental factors like scattering with substrate conduction electrons and phonons~\cite{miyamachi2013stabilizing,gatteschi2003quantum,otte2008role,PhysRevLett.124.077204,PhysRevB.97.024412}. Achieving this stability demands substantial energy barriers between different spin states, meaning significant magnetocrystalline anisotropies. 

Foundational work has been laid through experimental studies investigating rare-earth atoms on graphene or metallic substrates \cite{forster2012phase,jugovac2023inducing,herman2022tailoring,PhysRevB.90.235437,PhysRevX.10.031054,PhysRevB.96.224418,PhysRevLett.113.237201,donati2021correlation}, employing techniques such as x-ray adsorption spectroscopy, x-ray magnetic circular dichroism, and scanning tunneling microscopy to discern magnetic ground states, excitation processes, and relaxation times \cite{singha2021mapping,baltic2018magnetic,baltic2016superlattice,bellini2022slow,PhysRevLett.130.106702}, and showing that achieving stable magnetic units in the single-atom limit is possible in some instances, as demonstrated for single Ho atoms adsorbed on MgO \cite{donati2016magnetic}. 
\begin{figure*}[t!]
    \includegraphics[width=\textwidth]{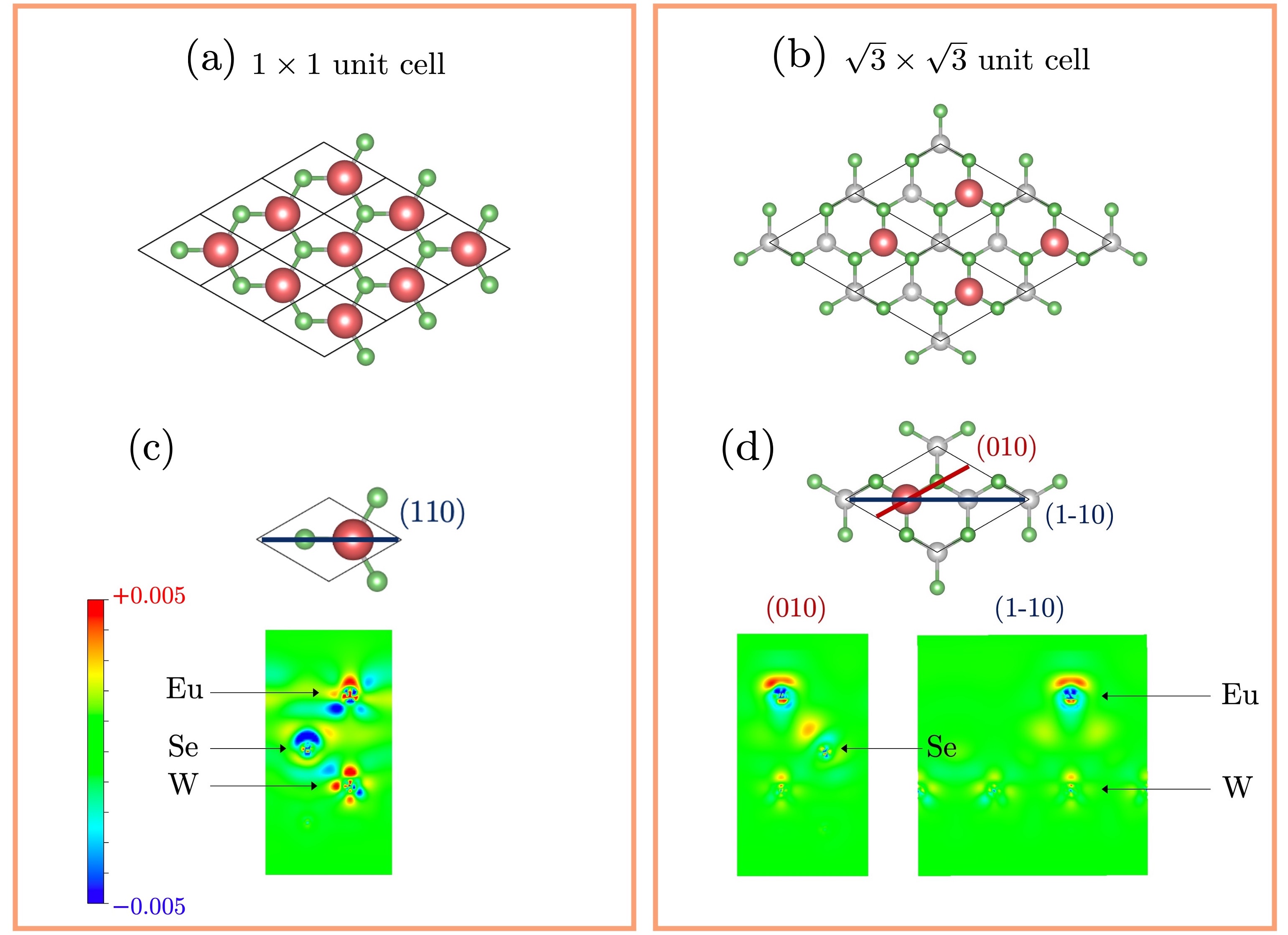}
    \caption{Rare-earth atoms (depicted as red spheres) adsorbed on the T-W site on the W atoms (grey spheres) of a 1H-WSe$_2$ monolayer: (a) $1\times1$ unit cell, representing a high coverage scenario, and (b) $\sqrt{3}\times\sqrt{3}$ supercell, illustrating a low coverage scenario of RE atoms. Se atoms are shown in green. The view is from the top. (c) and (d) show the corresponding differential charge densities in $e/\text{bohr}^3$ along different crystallographic planes, where blue regions denote a loss of charge, and red regions signify a gain in charge.}
    \label{Fig1}
\end{figure*}
However, predicting which rare-earth atom might yield a substantial magnetic anisotropy on a specific substrate is challenging due to variations in the lanthanide series' chemical nature and its interactions with the underlying material. To address this issue, we propose a theoretical study based on DFT+$U$ calculations. The objective is to compare the magnetic anisotropy of two different rare-earth representatives from a chemical perspective on a selected 2D material, a transition-metal dichalcogenide (TMDC) monolayer, and to examine variations in magnetic properties with the concentration of the rare-earth atoms adsorbed on the 2D material, providing insights into combinations of rare-earth atoms and coverage density that might be experimentally appealing for applications relying on magnetic stability.

In detail, we conduct a comparative study on the electronic and magnetic properties of three distinct RE atoms, Eu, Gd, and Ho, deposited at two different coverages on a monolayer of  1H-WSe$_2$, a valleytronic semiconductor, focusing on the ferromagnetic order. The two different adsorption density scenarios include one with a full monolayer coverage of RE atoms in a $1\times 1$ unit cell, equivalent to one magnetic atom per substrate unit cell, and the other with a diluted arrangement in a $\sqrt{3}\times\sqrt{3}$ supercell configuration, representing a 1/3 coverage, as depicted in Fig.~\ref{Fig1} (a) and (b), respectively.

The choice of RE atoms includes representatives with different electron configurations: On the one hand, Eu and Gd with half-filled $4f$ shells ($4f^7$), characterized by a small orbital angular momentum, and on the other hand, Ho with an open $4f$ shell ($4f^{11}$), which exhibits strong spin-orbit coupling (SOC) effects due to its non-zero orbital angular momentum. This variation in electronic structure affects the shape of the $4f$ charge cloud \cite{skomski2020anisotropy}, leading to different interactions with the $C_{3v}$ crystal field of the substrate, depending on the direction of magnetization. Another aspect of the choice of RE atoms are their valencies. Gd is considered as a special case due to its unique atomic valence $d^1$ configuration in $4f^75d^1$ having an additional outer $5d$ electron, which results in distinct chemical behaviors compared to Eu  and Ho, which have typically an atomic configuration of $4f^n5d^0$  with $n=7, 11$, respectively, lacking the outer $5d$ electron.  Consequently, this investigation sheds light on how to manipulate magnetocrystalline anisotropy by exploiting the chemical properties of the RE atom and its coverage on the 2D material. 

The choice of the TMDC 1H-WSe$_2$ as 2D material finds justifications in its inherent characteristics. This includes its robust SOC induced by the tungsten W atoms and its chemical composition, which is anticipated to result in strong orbital hybridizations with the adsorbed RE atoms.
This contrasts with previous theoretical studies  \cite{PhysRevB.82.245408,PhysRevB.94.125113,shick2019magnetism,PhysRevB.102.064402,,basiuk2022adsorption,carbone2023magnetic} that focused on the magnetic properties of RE elements on a graphene sheet, where the interaction process involved only the $\pi$ orbitals of the 2D material, and spin-orbit coupling would arise solely from the RE element. Moreover, the semiconducting 1H phase of WSe$_2$ is expected to exhibit a low density of states around the Fermi energy, potentially reducing scattering events between the magnetization of the RE atoms and conduction electrons, which is advantageous for the investigation of the quantum spin behavior of RE atoms at low temperatures. 

Limited information is available regarding the magnetic characteristics of RE atoms on TMDCs, as previous studies have primarily concentrated on introducing RE defects into TMDC monolayers to modify their electrical properties \cite{li2021enhanced,zhao2023engineering}. In contrast, our investigation delves into the adsorption of magnetic RE atoms onto these 2D materials, exploring the interactions within this heterostructure and their impact on magnetocrystalline anisotropy.

\section*{Adsorption properties}
\setlabel{Adsorption geometry and properties}{Ads}

\begin{table*}[t!]
\centering
\arrayrulecolor{myorange} 
%{\linewidth}  @{\extracolsep{\fill}}
\begin{tabular}{c!{\color{myorange!30}\vrule}ccc!{\color{myorange!30}\vrule}ccc}%{lllllll}
\midrule[0.4mm]
\multicolumn{2}{l}{ } &
\multicolumn{1}{l}{\textbf{$1\times1$}} & 
\multicolumn{1}{l}{} & 
\multicolumn{1}{l}{} &
\multicolumn{1}{l}{\textbf{$\sqrt{3}\times\sqrt{3}$}} &\\
%& & $1\times1$ & & & $\sqrt{3}\times\sqrt{3}$ & \\
\midrule[0.4mm]
Site &E$_\text{ads}$[eV] & $d_0$ [\r{A}]  & $d_\text{occ}$ & E$_\text{ads}$[eV] & $d_0$ [\r{A}]  & $d_\text{occ}$\\

%\midrule[0.2mm]
\rowcolor{myorange!15}
{\textbf{Eu atom}} & & & &  & &  \\

\text{T-W} &$-0.474$ &$2.500$ &$0.522$ &$-0.690$ &$2.500$ & $0.158$\\
\text{T-Se} &$-0.341$ &$3.119$ &$0.550$ &$-0.401$ &$3.112$ & $0.085$\\
\text{H}&$-0.312$ &$2.830$ &$0.520$ &$-0.611$ & $2.582$& $0.128$\\

\rowcolor{myorange!15}
{\textbf{Gd atom}} & & & &  & &  \\

\text{T-W} & $-0.507$ &$2.377$ & $1.020$&$-1.451$&$2.049$& $0.831$ \\
\text{T-Se} &$-0.415$ &$3.001$ &$1.000$ &$-0.688$ & $2.796$ & $0.684$\\
\text{H}&$-0.319$ &$2.716$ &$1.011$ &$-1.106$ &$2.162$ & $0.756$\\

\rowcolor{myorange!15}
{\textbf{Ho atom}} & & & &  & &  \\
\text{T-W} &$-0.946$ &$2.404$ &$0.440$ & $-0.634$ &$2.428$&$0.141$ \\
\text{T-Se} &$-0.780$ &$3.101$ &$0.405$ &$-0.382$ &$3.120$&$0.056$ \\
\text{H}&  $-0.511$&$2.672$ &$0.395$ &$-0.374$& $3.069$&$0.039$\\

\midrule[0.4mm]
\end{tabular}
\caption{\label{table1} Adsorption properties for Eu, Gd and Ho on a 1H-WSe$_2$ monolayer located at the T-W, T-Se and H sites for the $1\times 1$ and $\sqrt{3}\times\sqrt{3}$ coverages: adsorption energy in eV, adsorption distance in \AA{} (taken as the perpendicular distance from the first Se layer), and $d$ occupation of
the magnetic RE atom measured in the muffin-tin sphere with radius of 2.8\ $a_0$. Calculations for Eu and Gd have been performed without SOC. Calculation for Ho have been performed with SOC.}
\end{table*}

We begin the discussion by examining the adsorption features of the three rare-earth atoms on the 1H-WSe$_2$ monolayer. A concise discussion on the choice of coverage density in the Eu/WSe$_2$ system is available in Ref~\cite{carbone2022engineering}, exploring the potential for engineering anomalous Hall conductivity. On the WSe$_2$ substrate, we explore three distinct adsorption sites for the rare-earth atoms: atop a W atom (T-W), atop a Se atom (T-Se), or within the hexagonal \textquote{Hollow} site (H). To determine the adsorption energy, we subtract the total energy of the heterostructure, denoted as $E_{\text{RE/WSe}_2}$, from the total energies of the isolated components, represented by $E_{\text{RE}}$ and $E_{\text{WSe}_2}$,
\begin{equation}
    E_\text{ads}=E_{\text{RE/WSe}_2}-E_{\text{RE}}-E_{\text{WSe}_2}.
\end{equation}
Table~\ref{table1} presents the calculated adsorption properties in the $1\times 1$ and $\sqrt{3}\times\sqrt{3}$ simulation cells, for all three positions. Comparing the adsorption energies of different adsorption sites, it is evident that the preferred position for the RE atoms is atop the W atom in both instances. The adsorption of the RE element on WSe$_2$ induces a $C_{3v}$ symmetry at the local position. In addition, the RE atom at the T-W site is closer to the TMDC substrate compared to the T-Se and H sites, likely resulting in a more pronounced crystal field effect and charge transfer.

Starting with the discussion on Eu, in the diluted $\sqrt{3}\times\sqrt{3}$ configuration, the observed larger distances $d_0$ at the T-Se and H-sites, compared to the T-W site, correlate with a systematic decrease in the $5d$ occupation of the Eu atom from T-W to H to T-Se. This reduction in $d$ occupation primarily results from decreased interactions with the underlying substrate, and the smaller the interaction, the closer one comes to the atomic divalent limit with the $d^0$ configuration.
Conversely, in the $1\times1$ unit cell, the $d_\text{occ}$ values remain similar across adsorption sites, indicating that they predominantly originate from hybridization with neighboring Eu atoms in the magnetic monolayer and are not influenced by the distance from the substrate. Moreover, by comparison of the equilibrium distances between the two coverages, it is seen that the magnetic dilution only affects the perpendicular distance at the hexagonal H-site, which is reduced by $0.25$\ \AA.
Focusing on the T-W site, the most notable distinctions between the high coverage and low coverage scenarios are evident in the $d$ occupation. In the diluted limit, the occupation numbers approach the atomic configuration $d^0$ with a reduced $d$ occupation to $0.158$ electrons. This mirrors the contributions from neighboring Eu atoms and the WSe$_2$ monolayer discussed earlier. 

In Gd/WSe$_2$, the $1\times1$ configuration exhibits a larger $5d$ occupation as compared to the $\sqrt{3}\times\sqrt{3}$ scenario across all adsorption sites. Specifically, at the favored T-W position, it is larger by $0.19$ electrons, implying charge transfer from Gd $5d$ valence shell to the substrate in the dilute limit. This is further reflected by the increased adsorption energy observed when Gd atoms are more diluted, indicating notable hybridization effects involving these electrons. Intriguingly, this enhanced interaction with WSe$_2$ results in closer proximity, manifested in the considerably reduced equilibrium adsorption distance $d_0$. Precisely, $d_0$ measures $2.377$ \AA\ in the $1\times1$ coverage, lowered to $2.049$ \AA\ in the diluted T-W configuration. These data reveal a clear difference in chemical behavior compared to Eu, where dilution did not significantly affect the adsorption distance. 

Examining the adsorption characteristics of Ho atoms on WSe$_2$, we note a chemical behavior similar to that of Eu. Particularly, in the preferred T-W and T-Se sites, the equilibrium distance from the 2D material remains relatively similar upon dilution of the magnetic atom, while it undergoes changes in the H-site. Regarding the favored T-W adsorption site, the $d_\text{occ}$ is larger in the $1\times1$ cell compared to the $\sqrt{3}\times\sqrt{3}$ supercell, suggesting again an almost atomic configuration in the diluted scenario.

\section{Electronic properties}

\begin{table*}[t!]
\centering
\arrayrulecolor{myorange} 
%{\linewidth}  @{\extracolsep{\fill}}
\begin{tabular}{c!{\color{myorange!30}\vrule}ccc!{\color{myorange!30}\vrule}ccc}%{lllllll}
\midrule[0.4mm]
\multicolumn{2}{l}{ } &
\multicolumn{1}{l}{\textbf{$1\times1$}} & 
\multicolumn{1}{l}{} & 
\multicolumn{1}{l}{} &
\multicolumn{1}{l}{\textbf{$\sqrt{3}\times\sqrt{3}$}} &\\
%& & $1\times1$ & & & $\sqrt{3}\times\sqrt{3}$ & \\
\midrule[0.4mm]
 &$f_\text{occ}$ & $m_s^{\text{RE}}$[$\mu_\text{B}$] &$m_l^{\text{RE}}$[$\mu_\text{B}$] &$f_\text{occ}$ &$m_s^{\text{RE}}$[$\mu_\text{B}$] & $m_l^{\text{RE}}$[$\mu_\text{B}$] \\

%\midrule[0.2mm]
\rowcolor{myorange!15}
{\textbf{Eu atom}} & & & &  & &  \\
\text{T-W}& $6.865$ &$7.130$ & $-0.032$ & $6.923$ &$6.994$& $-0.005$\\
\text{T-Se} & $6.858$ &$7.440$ & - & $6.922$ & $7.000$ & - \\
\text{H}& $6.861$ & $7.240$& - &$6.923 $ &$6.991$ & -\\

\rowcolor{myorange!15}
{\textbf{Gd atom}} & & & & & &\\
\text{T-W}& $6.980$ & $7.509$ & $-0.023$ &$7.012$&$7.407$& $-0.166$  \\
\text{T-Se}&$6.977$ & $7.564$ & -&$6.984$&$7.530$& - \\
\text{H}&$6.980$ &$7.589$ & - &$7.004$ &$7.433$& -\\

\rowcolor{myorange!15}
{\textbf{Ho atom}} & & & &  & &  \\
\text{T-W$_{hunds}$} & - &-&-&$10.910$ &$3.012$&$5.935$ \\
\text{T-W} & $10.846$ &$3.068$&$4.888$&$10.910$ &$3.009$&$4.948$ \\
\text{T-Se}&$10.849$ &$3.107$&$4.899$&$10.911$ &$3.008$&$4.956$\\
\text{H}& $10.853$& $3.085$&$4.897$&$10.910$ &$3.005$&$5.944$\\
\midrule[0.4mm]
\end{tabular}
\caption{\label{table2} Ground state properties for Eu, Gd and Ho on a WSe$_2$ monolayer in the T-W, T-Se and H sites in the $1\times1$ and $\sqrt{3}\times\sqrt{3}$ coverages: $f$ occupation of the magnetic RE atom, spin magnetic moment of the RE atom in $\mu_{\text{B}}$, and orbital magnetic moment in $\mu_{\text{B}}$, 
 all measured in the muffin-tin sphere with radius of 2.8\ $a_0$. Calculations for Eu and Gd have been performed without SOC except for the orbital moment in the T-W site. Calculations for Ho have been performed with SOC. The SOC calculations were performed in the presence of a perpendicular magnetization to the WSe$_2$ monolayer.}
\end{table*}

After exploring the adsorption geometries, our focus shifts to the electronic properties of the magnetic 2D heterostructures. Table~\ref{table2} presents the $f$ occupation, spin ($m_s$), and orbital ($m_l$) magnetic moments of the RE atoms in the system, measured in Bohr magnetons. The $m_l$ value is specifically assessed for the favored adsorption site T-W in the case of half-filled $4f$ shells (Eu and Gd). In the case of Ho, we computed the orbital moment for all three positions for a comprehensive comparison.

Examining Eu/WSe$_2$, we observe a few general trends: Firstly, irrespective of the Eu density, the spin moments of the Eu atom is larger than the occupancy of the localized $4f$ electrons. Assuming full spin-polarization of the $4f$ electrons, the intra-atomic exchange polarizes the primarily $5d$ valence electrons ferromagnetically. Secondly, reducing the concentration of the Eu atoms, we notice that Eu becomes more atomic like, the $f$ occupancy increases and the $d$ occupancy decreases, the spin-moment results more from the $4f$ electrons and less from the $d$ electrons. Thirdly, We find a small induced orbital moment. The negative sign indicates that it couples  antiparallel to the spin moment. We will see below that Gd/WSe$_2$ exhibits similar behavior, whereas Ho shows different behavior due to its open $4f$ shell. 

Focusing on Eu in the T-W site, we observe that, akin to the earlier-discussed $d$ occupation, the $f$ occupation increasingly mirrors the atomic configuration $4f^7$ when the atom becomes more diluted, increasing the $f$ occupancy from $\sim 6.87$ to $\sim 6.92$.  The distinct atomic behavior is also evident by the fact that in case of the  $\sqrt{3}\times \sqrt{3}$ supercell, the total spin magnetic moment comes practically only from the $4f$ electrons, while the $6.865$ unpaired $4f$ electrons in the $1\times 1$ unit cell contribute to a value of $\sim 7.130$ $\mu_{\text{B}}$. The additional $0.265$ $\mu_{\text{B}}$ results from the intra-atomic spin-polarization of the acquired $5d$ electrons. Conversely, in the $\sqrt{3}\times\sqrt{3}$ system, where the acquired $d$ occupation is only $0.158$ electrons, the spin magnetic moment of the RE is larger by merely $0.07$ $\mu_{\text{B}}$ compared to the pure $4f$ magnetic moment.  Additionally, the orbital magnetic moment undergoes a reduction by one order of magnitude upon dilution of the Eu atom.

As demonstrated in Ref.~\cite{carbone2022engineering}, diluting the magnetic Eu atom results in a reduction of magnetic proximity, manifested by a reduced exchange splitting of the energy states. This is qualitatively evident in the differential charge densities depicted in Fig.~\ref{Fig1}(c) and (d), calculated as $\Delta\rho(r)=\rho_{\text{Eu/WSe$2$}}(r)-\rho_{\text{Eu}}(r)-\rho_{\text{WSe$_2$}}(r)$. Red regions indicate a gain of charge, while blue regions signify a loss of charge. The more intense color spots suggest greater charge involvement in the formation of the heterostructure in the magnetic monolayer case, indicating stronger orbital hybridizations and magnetic proximity in the system.
\begin{figure*}[t!]
    \centering
    \includegraphics[width=\textwidth]{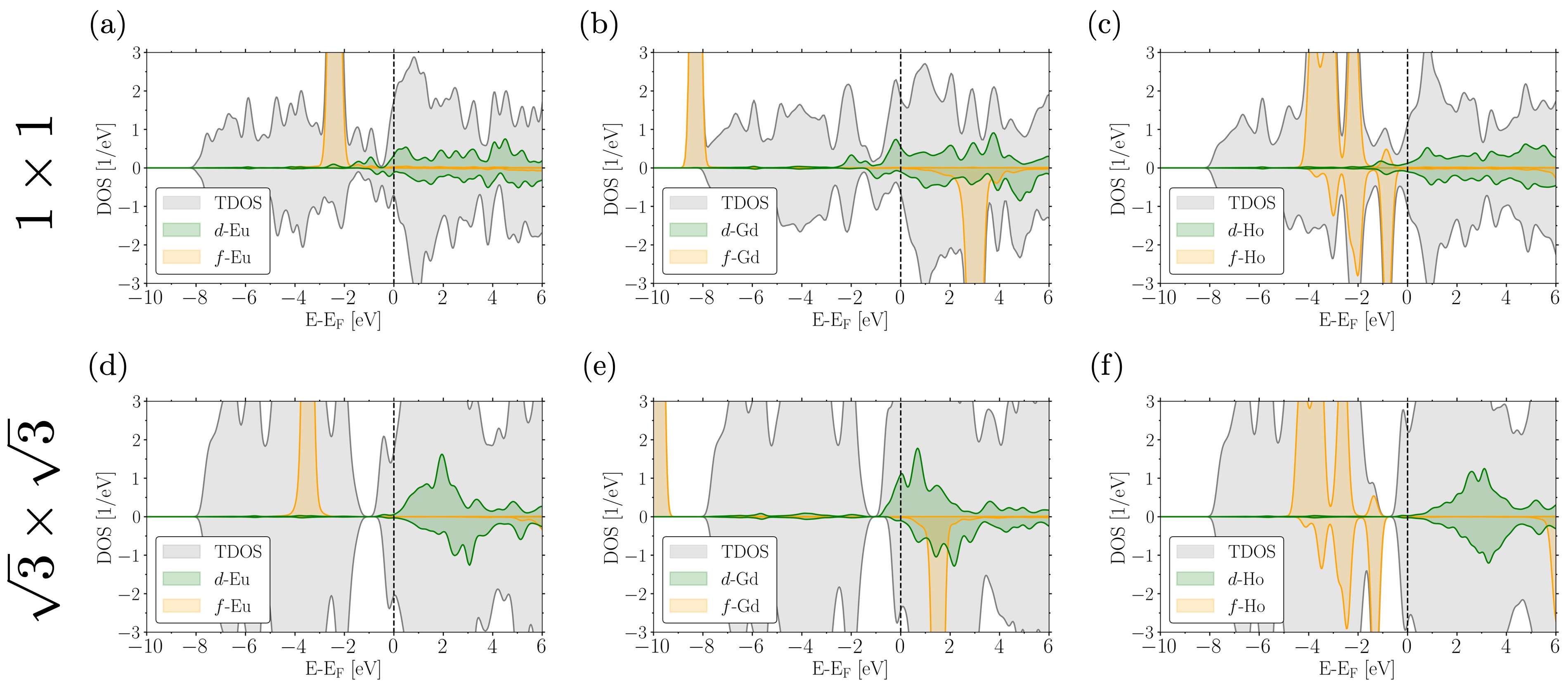}
            \caption{(a), (b), (c) Spin-polarized density of states for Eu, Gd, and Ho adsorbed on WSe$_2$ in $1\times 1$ unit cells. (d), (e), (f) Spin-polarized density of states for Eu, Gd, and Ho adsorbed on WSe$_2$ in $\sqrt{3}\times \sqrt{3}$ supercells. The shaded curves represent the total density of states of the heterostructures, while orange and green denote the $4f$ and $5d$ states, respectively, of the RE atoms. The upper half of the plots corresponds to the spin-up channel, and the lower half to the spin-down channel. }
    \label{Figure2}
\end{figure*}

Gd, on the other hand, exhibits a significantly larger overall spin magnetic moment, with values of $7.509$ $\mu_\text{B}$ in the $1\times1$ unit cell and $7.407$ $\mu_\text{B}$ in the $\sqrt{3}\times\sqrt{3}$ supercell, compared to $7.130$ $\mu_\text{B}$ and $6.994$ $\mu_\text{B}$ for the Eu case at the favored T-W site. These increased moments are attributed to the higher $5d$ occupation, which is intra-atomically spin-polarized by the $4f$ magnetic moment.
This deviation is also evident in the change of orbital magnetic moment, which increases in magnitude from $-0.023$ $\mu_\text{B}$ to $-0.166$ $\mu_\text{B}$ when diluting the Gd atom, in contrast to the decreasing trend observed for Eu from $-0.032$ $\mu_\text{B}$ to $-0.005$ $\mu_\text{B}$.

Regarding the open $4f$-shell system of Ho, in the $1\times 1$ unit cell the $f_\text{occ}$, along with the spin and orbital magnetic moments, reveals that Ho essentially retains its atomic $4f$ occupation with 11 electrons, three of which are unpaired. However, in the T-W site the orbital occupation deviates from Hund's rules by 1 $\mu_{\text{B}}$, which leads to a value of $5\ \mu_{\text{B}}$ for $m_l^{\text{Ho}}$. Consequently, the adsorption on top of WSe$_2$ induces quenching of the magnetic orbital moment driven by the competition between the crystal field and the intra-orbital exchange interaction. This phenomenon is also observed in the other adsorption sites, leading to values of $m_l\sim 4.9$ $\mu_{\text{B}}$. The observed quenching can be attributed to both crystal field effects from the substrate and the closely lying Ho nearest neighbors in the magnetic monolayer.

In the diluted $\sqrt{3}\times\sqrt{3}$ case, simulations on the T-W site lead to two different energy minima: one following Hund's rules with $m_l^{\text{Ho}}\sim6 \ \mu_{\text{B}}$, and another with a quenched value of $m_l^{\text{Ho}}\sim5 \ \mu_{\text{B}}$. Similarly to the $1\times1$ case, in the T-Se site, the orbital moment is again quenched, while it follows Hund's rules in the H-site.
These behaviors can be understood by considering the hybridizations with the surrounding crystal field, which appear stronger when the adatom adsorbs directly on top of an atom of the WSe$_2$ monolayer. Conversely, the effect is reduced when adsorbing in the middle of the hexagon formed by Se and underlying W atoms. It is important to note that these calculations of Ho atoms have been performed in the presence of self-consistently included SOC effects by choosing a perpendicular spin-quantization axis (along $z$) to the substrate. Additionally, in the T-W site, the adatom with a $4f$ orbital occupation deviating from Hund's rules has not been further relaxed compared to the T-W$_{hunds}$ case.

To examine the distinctions of the electronic structure among the investigated rare-earth systems, we refer to Fig.~\ref{Figure2}, which illustrates the spin-resolved density of states (DOS) for each chosen rare-earth atom under the two concentration scenarios.
Specifically, Fig.~\ref{Figure2}(a), (b), and (c) correspond to the $1\times 1$ magnetic monolayer case, while Fig.~\ref{Figure2}(d), (e), and (f) pertain to the diluted situation in the $\sqrt{3}\times \sqrt{3}$ supercell.

Notably, both Eu and Ho exhibit a similar behavior, with the occupied $4f$ states (orange peaks) positioned closer to the Fermi energy compared to Gd. This holds true for both coverages, especially in the $1\times 1$ unit cell, where the $4f$ peaks tend to come in close energetic proximity with other states of the heterostructure, such as the $d$ states of the RE atom itself.
\begin{figure}[b!]
    %\centering
    \includegraphics[scale=0.105]{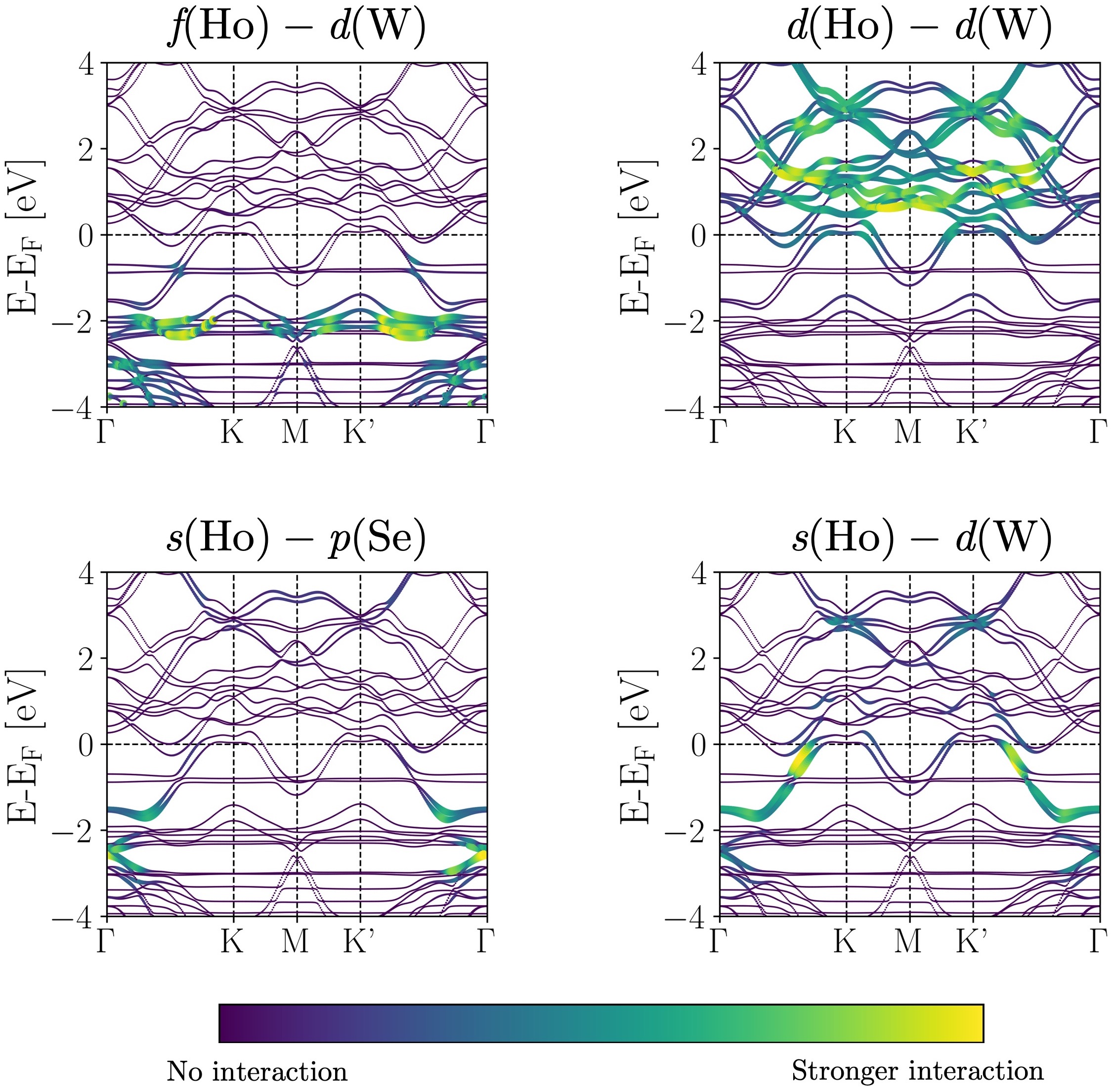}
%[width=0.7\textwidth]
    \caption{SOC-included calculations of orbital interactions in $1\times 1$ Ho/WSe$_2$ along the $\Gamma$-K-M-K'-$\Gamma$ path. The interactions include: $f$ states of Ho with $d$ states of W, $d$ states of Ho with $d$ states of W, $s$ states of Ho with $p$ states of
Se, and $s$ states of Ho with $d$ states of W. }
    \label{Figure3}
\end{figure}
Particularly, Ho exhibits a splitting of the $4f$ manifold attributable to the influence of SOC and electron correlation, thus covering a broader energy range.  This proximity facilitates orbital hybridization, as qualitatively shown in Fig.~\ref{Figure3}, which illustrates the product of the weights of different orbitals projected to various combinations of chemical species in $1\times 1$ Ho/WSe$_2$. The remarkably flat $4f$ bands are observable within an energy span of $\sim -4$ eV to $\sim -0.8$ eV. Additionally, the distinctive K-valleys of the TMDC manifest around $\sim -1.8$ eV at the high-symmetry points K and K', with an energy splitting induced by spin-orbit coupling. Similar to the findings in Ref.~\cite{carbone2022engineering} for Eu on WSe$_2$, we observe a direct interaction between the $4f$ electrons and the delocalized electrons constituting the crystal field, \textit{e.g.}\ $f$ electrons of Ho with $d$ electrons of W, along with interactions between the spin-polarized delocalized electrons of the system, such as between $d$ of Ho and $d$ of W. 

Examining the DOS for Gd atoms on WSe$_2$ (Fig.~\ref{Figure2}(b) and (e)), the scenario is notably different. The occupied $4f$ peaks are deep in energy ($\sim -8$ eV for $1\times1$ and $\sim -10$ eV for $\sqrt{3}\times\sqrt{3}$). This considerable energy separation prevents orbital hybridization since these peaks are energetically far removed from all other states in the system. However, the spin-polarized $d$ states exhibit a significant enhancement at the Fermi energy compared to Eu and Ho, particularly in the $\sqrt{3}\times\sqrt{3}$ diluted case.

\section{Magnetic anisotropy}
\begin{figure*}[t!]
    \centering
    \includegraphics[width=\textwidth]{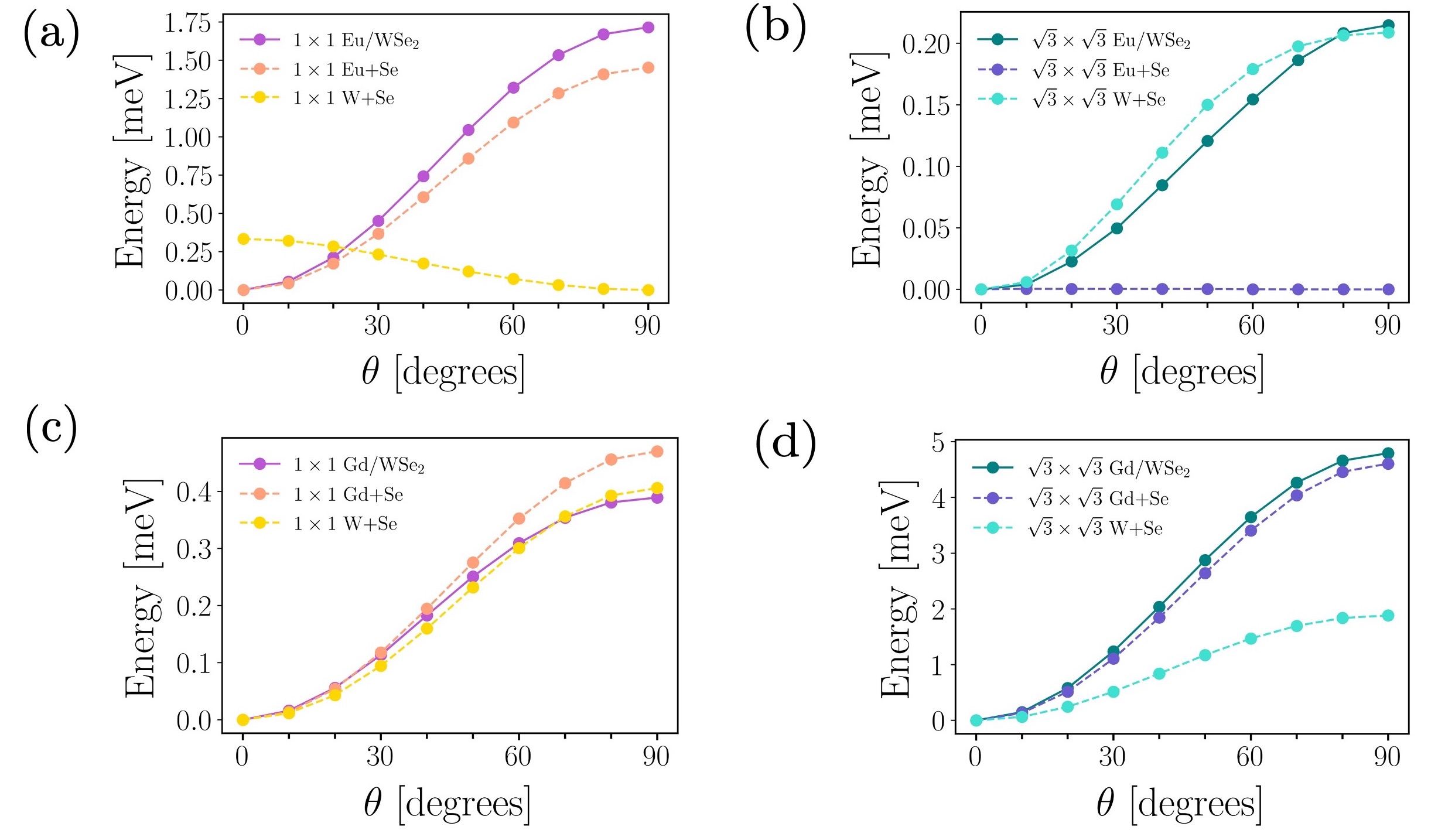}
    \caption{(a) Magnetic anisotropy energy curves are depicted for the $1\times1$ magnetic monolayer of Eu/WSe$_2$ (solid purple line) in (a), showcasing the distinct contributions from Eu+Se (dotted orange) and W+Se (dotted yellow) components. In (b), the MAE of the diluted $\sqrt{3}\times\sqrt{3}$ configuration of the overall Eu/WSe$_2$ system is presented (solid dark green line), along with the individual contributions from Eu+Se (dotted blue) and W+Se (dotted turquoise) components. Subsequently, (c) and (d) showcase the corresponding MAE curves for the $1\times1$ and $\sqrt{3}\times\sqrt{3}$ Gd/WSe$_2$, along with the contributions from the respective components. The total energy of the system is plotted against the polar angle $\theta$ of the magnetization, measured from the $z$-axis.}
    \label{Figure4}
\end{figure*}
The impact of magnetic dilution on the magnetocrystalline anisotropy energy (MAE) is shown in Fig.~\ref{Figure4}, depicting the total energy variation as a function of the magnetization orientation for Gd and Eu on WSe$_2$ in the two coverage scenarios. Each plot illustrates the MAE for the entire magnetic heterostructure using solid lines, along with the corresponding contributions from the individual components, RE+Se and the 2D material W+Se,  represented by dotted lines and obtained by switching off sequentially the SOC contribution at the W atoms and at the RE atom. Figures \ref{Figure4}(a) and (b) correspond to $1\times1$ and $\sqrt{3}\times\sqrt{3}$ Eu/WSe$_2$, while Figures \ref{Figure4}(c) and (d) present the analogous data for Gd/WSe$_2$.
Specifically, the magnetization rotates from perpendicular to the WSe$_2$ plane (along the $z$-axis, $\theta=0^{\circ}$) to parallel to the WSe$_2$ plane (along the $x$-axis, $\theta=90^{\circ}$).  All energy values are shifted with respect to the lowest energy point, set at $0$ eV.

The overall anisotropic behavior for both RE atoms exhibits a low-order nature, primarily governed by the first magnetic anisotropy constant, denoted as $K_1$ in $E_{an}=K_1\sin^2\theta$, where $\theta$ represents the polar angle associated with out-of-plane anisotropy. For both magnetic systems and coverages, the system favors an out-of-plane magnetization direction, as evidenced by the lowest energy at $\theta=0^{\circ}$ in the solid lines in purple and dark green.

Specifically for Eu/WSe$_2$, in comparing the magnetic anisotropy between the $1\times1$ and $\sqrt{3}\times\sqrt{3}$ cells (Fig.~\ref{Figure4}(a) and (b)), we observe that the energy barrier required for magnetization rotation from perpendicular to parallel to the 2D-material is approximately $1.75$ meV for the $1\times1$ cell and roughly $0.20$ meV for the $\sqrt{3}\times\sqrt{3}$ cell. This suggests the need for a high coverage of Eu atoms to achieve stable magnetic states. 
The difference can be explained in terms of orbital hybridizations. In the $1\times1$ high coverage scenario, the close proximity of the spin-up occupied $4f$ states to Eu $5d$ states and other delocalized substrate states (\textit{e.g.}\ $d$ states of W) leads to orbital hybridization (see Fig.~\ref{Figure2}). Thus, the pronounced anisotropy can be fundamentally attributed to two main factors:

1) Direct interaction: an interaction between the $4f$ charge and other delocalized electrons of WSe$_2$, such as the $d$ states of W. This interaction induces orbital angular momentum in the $4f$ shell, resulting in an anisotropic interaction with the surrounding crystal field upon rotation of the magnetization.
2) Indirect contribution: this is driven by intra-atomic spin-polarization from the $4f$ states to the $d$ states of Eu. The latter, being more spatially extended, interact anisotropically with the environment based on the rotation of the $4f$ magnetic moment through intra-atomic exchange interaction. However, given the small $5d$ occupation, this contribution is expected to have a lesser influence on the overall energy scale.

In the diluted $\sqrt{3}\times\sqrt{3}$ case, the factors contributing to magnetic anisotropy are expected to be analogous but less pronounced due to the lower density of magnetic atoms per W atom. The W atoms act also as the origin of spin-orbit coupling, essentially \textquote{inducing} SOC in the magnetic source. Consequently, the interplay between magnetism and SOC results in a less prominent outcome compared to the high coverage scenario.

Upon examining the contributions from Eu+Se and W+Se in the two coverage scenarios by switching on SOC only in the MT spheres of specific atoms, it becomes apparent that in the $1\times 1$ case, the predominant contribution arises from Eu atoms. Their spin-orbit contribution favors an out-of-plane easy axis, closely resembling the anisotropy energy barrier observed in the total system. In contrast, the TMDC alone favors an in-plane magnetization orientation (indicated by the yellow dotted line). With dilution, the Eu+Se contribution vanishes, revealing that all MAE contribution is attributed to the slightly spin-polarized WSe$_2$. Intriguingly, magnetic dilution induces a shift in the easy-axis of the magnetic TMDC.

Shifting our attention to Gd/WSe$_2$ (refer to Fig.~\ref{Figure4}(c) and (d)), which trend is reversed, suggesting that the diluted situation yields larger magnitudes compared to the high-coverage scenario. Specifically, for the $\sqrt{3}\times\sqrt{3}$ scenario, an energy difference of approximately $\sim 5$ meV is observed from the perpendicular to the in-plane direction, while in the $1\times 1$ unit cell, this is reduced to about $\sim 0.2$ meV. 

To explain the reduced MAE in the high-coverage scenario compared to Eu, it must be noted that the direct contribution is absent, and only indirect contributions are observed, as the $4f$ states are energetically too distant to interact with other states. While this holds true for the diluted scenario of Gd on WSe$_2$, the computed magnetic anisotropy remains remarkably large. In this case, the effect likely originates from the extensive spin-polarized $d$ DOS observed at the Fermi energy, coupled with the adsorption characteristics that result, as observed previously, in a very short equilibrium distance compared to all other cases. Therefore, the magnetic anisotropy in this case arises from the robust interaction of the spin-polarized $d$ states with the surrounding $C_{3v}$ crystal field.

Analyzing the individual contributions to the MAE, we observe that in the $1\times1$ scenario, both Gd and the spin-polarized substrate contribute similarly in magnitude and preferring an out-of-plane magnetization direction. In the diluted case, the primary contribution comes from the Gd atoms, although the effect of WSe$_2$ increases significantly by one order of magnitude due to its closer proximity to the Gd atom and, consequently, its stronger spin-polarization. 

In both Eu/WSe$_2$ and Gd/WSe$2$, the variation in the magnitude of MAE with magnetic dilution can also be explained by considering the change in the magnetic orbital moment of the RE atoms. The larger MAE of Eu/WSe$_2$ in the $1\times1$ unit cell corresponds to a larger value of $m_l$ by one order of magnitude. Similarly, the increased MAE in Gd/WSe$2$ from $0.4$ meV to approximately $5$ meV, as coverage is reduced, reflects a change in $m_l$ from $-0.023$ $\mu_{\text{B}}$ to $-0.166$ $\mu_{\text{B}}$. More details on the orbital moment anisotropy of the two RE/WSe$_2$ systems in the two coverage scenarios are available in the Supplementary Material. This includes the variation of the magnetic orbital moment relative to the magnetization direction.

\subsection{Magnetic anisotropy of Ho atoms on a WSe$_2$ monolayer}

\begin{figure}[b!]
    \centering
    \includegraphics[scale=0.55]{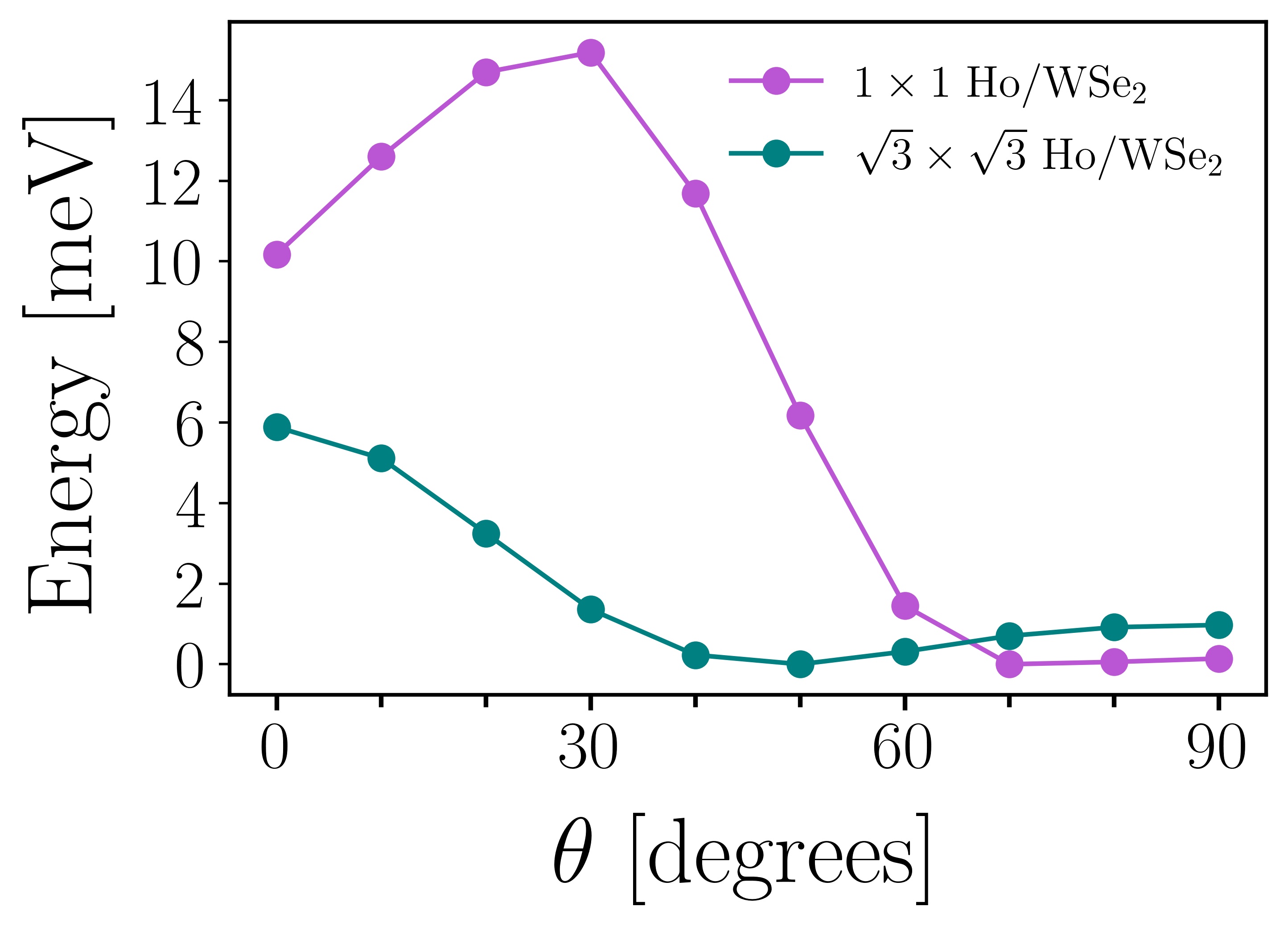}
    \caption{(a) Magnetic anisotropy energy curves for the $1\times1$ magnetic monolayer (purple) and the diluted $\sqrt{3}\times\sqrt{3}$ scenario (dark green) of Ho/WSe$_2$. The total energy of the system is plotted versus the polar angle $\theta$ of the magnetization, measured from the $z$-axis. }
    \label{Figure5}
\end{figure}

\begin{table*}[t!]
\centering
\begin{tabular*}{\linewidth}{@{\extracolsep{\fill}}
    l*{6}{S[table-format=-1.3]}}
          % {l *{5}{S[table-format=-1.4]}}%{@{}lllll}
\midrule[0.4mm]
\mc{\text{Cell}} &\mc{$K_{1}$} & \mc{$K_{2}$}& \mc{$K_{2}^{'}$} & \mc{$K_{3}$} & \mc{$K_5$}\\
\midrule[0.2mm]
 $1\times 1$&-56.061& -160.875&6.704& 94.143& -16.701\\ 
$\sqrt{3}\times \sqrt{3}$ &-28.095&37.341& 1.515&-14.189 & 3.950\\ 
\midrule[0.4mm]
\end{tabular*}
\caption{\label{table3}Magnetic anisotropy constants obtained via fitting of DFT+U data depicted
in Fig.~\ref{Figure5} (b) for $1\times 1$ and $\sqrt{3}\times\sqrt{3}$ Ho/WSe2. The values are reported in meV. }
\end{table*}
In examining the magnetic anisotropy of Ho/WSe$_2$, we observe that, in contrast to Gd, Ho follows a pattern similar to Eu: larger MAEs are evident in the $1\times1$ unit cell compared to the $\sqrt{3}\times\sqrt{3}$ coverage. This is depicted in Fig.~\ref{Figure5}, where energy values peak at around 16 meV in the $1\times1$ cell (purple), covering a range from the canted easy-axis direction at approximately $\theta=70^{\circ}$ to the hard-axis found at $\sim \theta=30^{\circ}$. In the $\sqrt{3}\times\sqrt{3}$ supercell (dark green), there is an energy difference of about 6 meV between the easy ($\sim \theta=50^{\circ}$) and hard ($\theta=0^{\circ}$) axes.

Additionally, in this case, the open $4f$-shell induces a highly anisotropic behavior in space, resulting in energy curves that deviate from the first contribution that is proportional to $K_1\sin^2\theta$. Specifically, in a trigonal $C_{3v}$
crystal field, the DFT+$U$ calculations can be fitted using the equation \cite{skomski2020anisotropy}:
\begin{equation}
 \begin{split}   
 E_{an}=&K_1\sin^2\theta+K_2\sin^4\theta \\
 +&K_2^{'}\sin^3\theta\cos\theta\cos(3\varphi) \\
 +&K_3\sin^6\theta+K_4\sin^6\theta\cos(6\varphi) \\
 +&K_5\sin^3\theta\cos^3\theta\cos(3\varphi),   
 \end{split}  
\end{equation}

where the $K_i$ are the magnetic anisotropy constants, and $(\theta,\varphi)$ are the polar and azimuthal angle describing the magnetization direction.
Focusing solely on the out-of-plane contribution ($\varphi=0$), we need to consider the magnetic anisotropy constants $K_1, K_2, K_2^{'}, K_3,$ and $K_5$. The values obtained from the fitting procedure are presented in Table~\ref{table3}. It is noteworthy that substantial contributions come from all $K_i$ constants, highlighting the significantly anisotropic behavior compared to the Eu and Gd systems.

We can attribute the highly anisotropic behavior of the Ho atoms to the direct and indirect contributions observed earlier for Eu, along with the influence of the open $4f$-shell that inherently induces substantial spin-orbit coupling effects. This effect is evident, for instance, in Ref.~\cite{carbone2023magnetic}, where large magnetic anisotropy values of Ho are found on a bare graphene monolayer that lacks SOC. In this scenario, the Ho atom resides within a high-symmetry hexagonal crystal field created by the C atoms, resulting in relatively weak orbital interactions. This allows for a straightforward application of a point-charge model to theoretically describe the magnetic states. The heightened complexity of magnetic anisotropy in the WSe$_2$ heterostructure is evident not only in the reduced symmetry leading to a larger number of magnetic anisotropy constants (for a $C_{6v}$ field, only $K_1$, $K_2$, $K_3$, and $K_4$ are relevant \cite{skomski2020anisotropy}) but also in the discussed orbital interactions due to the chemical variety in the chosen 2D material.  These interactions, while complicating the adoption of a point-charge model, can still be harnessed to achieve larger magnitudes in the MAE. 

From a quantum perspective, it is crucial to note that reducing the local symmetry around the magnetic source introduces additional quantum operators capable of mixing magnetic states. This potential mixing can favor quantum tunneling of magnetization \cite{gatteschi2003quantum,tejada1996quantum} and, consequently, destabilize the magnetization. For example, in a hexagonal $C_{6v}$ crystal field, the Stevens operator \cite{stevens1952matrix,hutchings1964point} $\hat{O}_6^6$ mixes states labeled by the quantum number $J_z$ in a total angular momentum manifold $J$ with a difference of $\Delta J_z=\pm 6$. On the other hand, a trigonal $C_{3v}$ crystal field includes extra operators that can generate quantum superpositions of states differing by $\Delta J_z=\pm 3$ in addition to the $\hat{O}_6^6$ operator. In principle, adsorbing the magnetic atom onto a high-symmetry site is preferred. However, if the mixed states do not represent the magnetic ground state, choosing a 2D material that creates substantial energy gaps between magnetic states (\textit{e.g.}, through orbital hybridizations) could lead to a favorable outcome for generating stable magnetic units.

Heading back to the comparison with  Eu, in Ho/WSe$_2$ it is crucial to consider that in the high-density $1\times1$ magnetic monolayer the compact $4f$ charge clouds themselves contribute to the crystal field, leading to a $4f$-$4f$ repulsion—a factor absent in Eu due to its spherical $4f$ charge cloud. Collectively, these factors contribute to an enhanced magnetic anisotropy compared to $1\times 1$ Eu/WSe$_2$.
\begin{figure*}[t!]
    \centering
    \includegraphics[width=\textwidth]{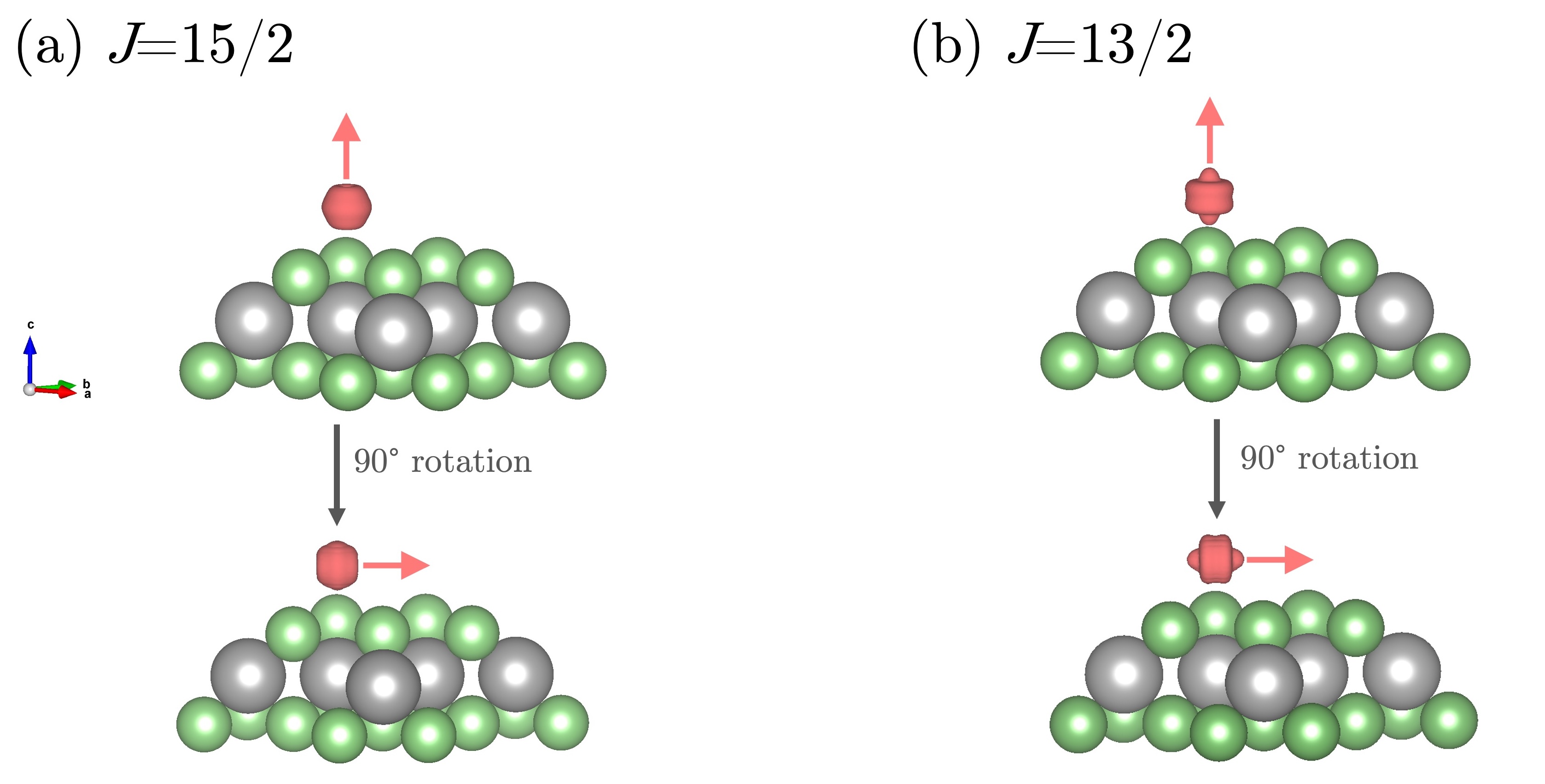}
      \caption{Magnetization densities of the Ho atom on top of WSe$_2$ in the two distinct orbital occupation configurations with perpendicular ($\theta=0^{\circ}$) and parallel ($\theta=90^{\circ}$) magnetization to the 2D-material. The isosurface level is $0.01$\ $e/\text{bohr}^3$. The red arrow symbolizes the spin quantization axis.}
    \label{Figure6}
\end{figure*}
Nonetheless, the pattern seems analogous, as the magnetic anisotropy energy decreases with the dilution of Ho atoms. In both Eu and Ho cases, this reduction can be attributed to a decreased magnetic proximity and interaction with the overall crystal field—both from the substrate and neighboring RE atoms. Another noticeable factor is the diminishing spin-polarized $d$ occupation near the Fermi energy, signifying a reduction in the indirect contribution.

Table~\ref{table2} for Ho indicates that for the $\sqrt{3}\times\sqrt{3}$ diluted coverage in the preferred adsorption site, T-W, the energy minimization occurs with two distinct $4f$ orbital occupations. One adheres to Hund's rules, exhibiting a magnetic orbital moment of $m_l^{\text{Ho}}=6$ $\mu_\text{B}$, signifying 7 electrons occupying orbitals with magnetic quantum numbers from $-3$ to $+3$ with aligned spins to maximize the total spin moment, while the remaining 4 electrons then fill the orbitals with magnetic quantum numbers +3, +2, +1, and 0, maximizing the total angular moment. In contrast, the configuration with $m_l^{\text{Ho}}=5$ $\mu_\text{B}$ involves displacing one spin-down electron from the orbital with magnetic quantum number 0 to $-$1, thereby quenching the orbital magnetic moment by 1\ $\mu_\text{B}$. In the case of Hund's rules orbital occupation, the total angular momentum is $J=15/2$, whereas for the deviation, it is $J=13/2$. As previously explained, this reduction in the orbital moment stems from a competition between the crystal field interaction and the intra-atomic exchange, indicating that the crystal field effect surpasses the latter in energy. This behavior contrasts with what is observed for Ho atoms on a graphene monolayer, as discussed in Ref.~\cite{carbone2023magnetic}, where the $4f$ orbital occupation adheres to Hund's rules.

To assess the impact of the quenching of the orbital magnetic moment on magnetic anisotropy, DFT+$U$ calculations were employed to compute the total energy for different out-of-plane magnetization directions, ranging from $\theta=0^{\circ}$ to $\theta=90^{\circ}$ for the magnetic state $J=13/2$. For angles between $0^{\circ}$ and $20^{\circ}$, the $J=13/2$ state is energetically favored, being 0.33 eV lower in energy compared to $J=15/2$. However, as the magnetization rotates further, an energy inversion occurs, and the $J=15/2$ state becomes the new ground state. Due to difficulties in achieving convergence for the $J=13/2$ state at larger angles, the self-consistent procedure was performed by fixing the $4f$ occupation matrix to the desired value of $J$. Consequently, we refrain from discussing energy differences between the two magnetic states.

%To assess the impact of quenching of the orbital magnetic moment on magnetic anisotropy, DFT+$U$ calculations were employed to compute the total energy for different out-of-plane magnetization directions, represented by the angle $\theta$ in Fig.~\ref{Figure6}. For perpendicular magnetization ($\theta=0^{\circ}$) along the $z$-axis, the magnetic state $J=13/2$ is the energetically favored ground state, being 0.33 eV lower in energy. At $\theta=10^{\circ}$, the $J=13/2$ state remains the magnetic ground state with a reduced energy difference of 0.11 eV from $J=15/2$. As $\theta$ increases to $\theta=20^{\circ}$, this energy difference further diminishes. However, starting from $\theta=30^{\circ}$, an energy inversion takes place, and the $J=15/2$ state becomes the new ground state. Subsequently, the magnetic anisotropy curve of $J=13/2$ exhibits substantial energy values, indicating its instability. 

%Nevertheless, our calculations reveal that the MAE depends on different contributions based on the $4f$ orbital occupation. For $J=15/2$, we observed that the canted easy-axis results from significant contributions of high-order magnetic anisotropy constants. Conversely, the $J=13/2$ scenario exhibits a magnetic anisotropy similar to the half-filled $4f$ shell cases, primarly dominated by the $K_1$ value.
Nevertheless, the consistent symmetry of the crystal field in both scenarios emphasizes that eventual differences in the MAE, both in shape and in magnitude, arise from the spatial geometry of the $4f$ charge cloud, determined by the varying $4f$ orbital occupation. This can be qualitatively observed in Fig.~\ref{Figure6}, where we show the computed magnetization densities of Ho within the two analyzed $4f$ orbital occupations in presence of a perpendicular spin quantization axis.
These calculations underline the necessity of an accurate description of the $4f$ states, as the magnetic properties can drastically change for different energy minima.

\section{Conclusion}
In this manuscript, we have demonstrated the tunability of magnetocrystalline anisotropy in rare-earth atoms on the valleytronic semiconducting 1H-WSe$_2$ monolayer based on their adsorption density. Specifically, the magnetic anisotropy is closely tied to the specific rare-earth atom chosen, and, in general, open $4f$-shells result in more significant energy variations as the magnetization is rotated in space. This is attributed to the non-spherical nature of the $4f$ charge density and the pronounced spin-orbit coupling arising from a non-zero orbital angular momentum.

We adopted density functional theory calculations to reveal that rare-earth atoms exhibiting chemical similarities to Eu, such as Ho without an external $5d$ valence electron, demonstrate more substantial magnetic anisotropies in high concentrations on the 2D material. Conversely, rare-earth atoms like Gd, possessing $5d$ valence electrons, exhibit the opposite behavior.

These contributions involve a direct interaction between localized $4f$ electrons and the environment, as well as an indirect contribution arising from intra-atomic spin-polarization and ferromagnetic exchange interaction between $4f$ electrons and delocalized electrons like $d$ electrons. These, in turn, interact with the crystal field. The strength of these interactions depends on coverage density, the proximity to the substrate, position of electronic states in the energy spectrum, and the geometry of the $4f$ charge density. These effects are intrinsic to the specific valence configuration of the rare-earth atom.

These findings bear experimental significance for applications requiring stable magnetizations, such as hard magnets, and they also encourage additional investigation into how the chemical properties of rare-earth atoms influence magnetic stability. Indeed, future analysis is required, considering various lanthanide species, both heavy and light (with more and fewer than half-filled $4f$ shells). Additionally, exploring the quantum-level impact on magnetic anisotropy, including determining the quantum multiplet splittings of magnetic states, would provide a more profound understanding of magnetization stability against quantum fluctuations. This analysis would also enable the eventual determination of two-level quantum systems. Moreover, such assessments should cover multiple options of transition metal dichalcogenide materials, involving variations in both metallic and chalcogen atoms within the substrate.

\section{Methods}
The results presented were obtained using the full-potential linearized augmented planewave (FLAPW) method as implemented in the FLEUR code \cite{FLEUR,wortmann2023fleur}. Reliable self-consistent results have been obtained adopting the PBE \cite{PBE} prescription of the generalized gradient approximation (GGA) \cite{PhysRevLett.77.3865} exchange-correlation functional with a $10\times 10$ $k$-point mesh, while the magnetic anisotropies were computed using
a $21\times 21$ $k$-point mesh.
The $1\times1$ unit cell, which included one RE atom, two Se atoms, and one
W atom, was simulated with the equilibrium GGA lattice constant of $a = 3.327$ \AA, as illustrated in Fig~\ref{Fig1}(a). Concerning the LAPW basis functions, the plane-wave basis cut-off was set to $K_{max}= 4.0\ a_0^{-1}$, and the maximum angular
momentum inside the MT spheres was set to $l_{max}= 10$ for the RE atom and $l_{max}=8$
for W and Se. To account for the highly localized $4f$ electrons of the magnetic atoms, the DFT+$U$ method was employed, considering the fully-localized limit for the double-counting term \cite{PhysRevB.60.10763}. The DFT+$U$ parameters were set to $U = 6.7$\ eV and $J = 0.7$\ eV for Eu and Gd, as per \cite{PhysRevB.60.10763,kurz2002magnetism}, and $U = 7.03$ \ eV and $J = 0.83$\ eV for Ho, based on reported values in \cite{shick2017magnetic}. It is noteworthy that for RE atoms, $U$ values around $7$ eV are widely accepted as they have been shown to reproduce experimental results for various properties of bulk RE systems \cite{PhysRevB.94.085137}.
For simulations of the dilute case, we utilize a supercell (Fig.~\ref{Fig1}(b)) with a lattice constant of $a=\sqrt{3}\times 3.327$ \AA\ and maintain the same self-consistent field parameters as in the $1\times 1$ cell, except for employing a $k$-point mesh of $20 \times 20$ for the spin-orbit coupling calculations. Calculations related to Ho atoms as well as those conducted for the magnetic anisotropy curves included the self-consistent treatment of the spin-orbit coupling by means of the second variation formulation \cite{PhysRevB.42.5433}.
\section{Code availability}
The underlying DFT code for this study is available in \cite{wortmann2023fleur} and can be accessed via this link \url{https://www.flapw.de/}. 
\section{Data Availability}
All data generated or analysed during this study are included in this published article and its supplementary information files. The datasets used and/or analysed during the current study available from the corresponding author on reasonable request.
\section*{Acknowledgments} 
The project is funded by the Deutsche Forschungsgemeinschaft (DFG) through CRC
1238, Control and Dynamics of Quantum Materials: Spin
orbit coupling, correlations, and topology (Project No.\ 277146847 C01). 
We acknowledge computing resources granted by RWTH Aachen University
under Project No.\ jara0219.
\section{Author Contributions}
J.P.C. ran the DFT calculations, post-processed the data, analyzed the results, and wrote the manuscript. G.B. assisted the data analysis and reviewed the manuscript. S.B. reviewed the manuscript and supervised the research. All authors equally discussed and conceived the research.
\section{Competing Interests}
All authors declare no financial or non-financial competing interests. 
%\bibliographystyle{unsrt}
%\bibliography{main.bib}
\printbibliography

\end{document}